\documentclass[review]{elsarticle}

\usepackage{amsmath,lineno,hyperref}

\journal{Journal of Magnetism and Magnetic Materials}

\begin{document}

\begin{frontmatter}

\title{Topologically Protected Dynamics of Spin Textures}


\author[mymainaddress]{O.~A.~Tretiakov\corref{mycorrespondingauthor}}
\cortext[mycorrespondingauthor]{Corresponding author}
\ead{olegt@imr.tohoku.ac.jp}
\author[mysecondaryaddress]{Ar.~Abanov}


\address[mymainaddress]{Institute for Materials Research, Tohoku University, Sendai 980-8577, Japan}
\address[mysecondaryaddress]{Department of Physics \& Astronomy,
	    Texas A\&M University,
            College Station, Texas 77843-4242, USA}

\begin{abstract}
We study current-induced dynamics of spin textures in thin magnetic nanowires. We derive effective equations of motion describing the dynamics of the domain-wall soft modes associated with topological defects. Because the magnetic domain walls are topological objects, these equations are universal and depend only on a few parameters.  We obtain spin spiral domain-wall structure in ferromagnetic wires with Dzyaloshinskii-Moriya interaction and critical current dependence on this interaction. We also find the most efficient way to move the  domain walls by resonant current pulses and propose a procedure to determine their dynamics by measuring the voltage induced by a moving  domain wall. Based on translationally non-invariant nanowires, we show how to make prospective magnetic memory nanodevices much more energy efficient.
\end{abstract}

\begin{keyword}
Domain walls dynamics, magnetic domain walls, ferromagnetic and antiferromagnetic
nanowires, topological spin textures
\end{keyword}

\end{frontmatter}


\section{Introduction}

Ferromagnets can be used to store and manipulate spin information, and new developments have created opportunities to use them as active components in spintronic devices \cite{Allwood01, Allwood02, Parkin08, Hayashi08}. Majority of the ideas to employ ferro- and antiferromagnets for memory or logic applications are related to the propagation of domain walls (DWs), skyrmions or other topological spin textures. This resulted in significant experimental \cite{Yamaguchi04,  Klaui:images05, Beach05,
  Thomas2006, HayashiPRL2006, Pi2011, Thomas2010} and theoretical \cite{Nakatani03, Zhang04, Tatara04, Thiaville05, Barnes05, Duine07, Tretiakov08, Clarke08, Tserkovnyak2008, Lucassen09, Duine09, Tretiakov_DMI, Min10, Tretiakov:losses, YanEPL10, Liu11,  Tretiakov:JAP, Brataas2011, Tretiakov2012, Tveten2013, Nagaosa2013} progress in this direction. To study theoretically the propagation of the spin textures usually numerical solutions of the Landau-Lifshitz-Gilbert (LLG) equation \cite{Zhang04, Thiaville05} are employed or the dynamics of softest modes for the motion of the topological textures is considered using collective coordinate approach. In this paper we overview the latter approach and consider several important cases of its application to the DW dynamics.  
  
\begin{figure}
\begin{center}
\includegraphics[width=0.45\columnwidth]{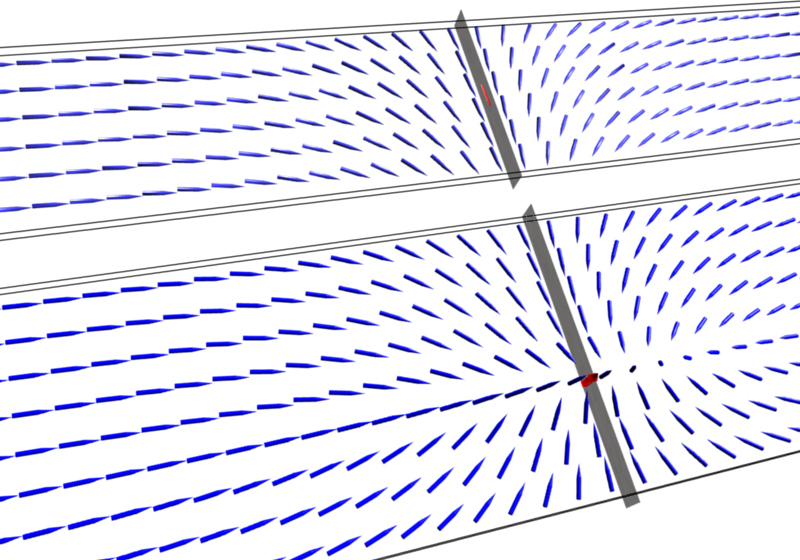}
\end{center}
\caption{Types of possible domain walls which can be described by two-component collective coordinate approach. The upper picture shows the simplest type of the domain wall: transverse DW.}
\label{fig:twoDWs}
\end{figure}  
  
As we will show, for the topologically robust spin textures the equations of motion for their lowest modes are universal and can be described by just a few parameters, which can be whether experimentally measured or analytically calculated for rather simple models. These softest modes are associated with the motion of topological defects comprising the DWs, see Fig.~\ref{fig:twoDWs}. Similar type of equations describe the dynamics of skyrmions \cite{Nagaosa2013} or composite skyrmions (vortex-antivortex pair with opposite magnetization polarizations at their core) \cite{Tretiakov07} in ferromagnets. Generally the lowest (zero) mode corresponds to the translation of the spin texture as a whole along the nanowire, which  is  thin  enough  to  have  a  homogeneous  magnetization  over  its
thickness. The other modes correspond to the texture rotations or the internal dynamics of the its topological defects. Below we mostly concentrate on the current driven DW dynamics, although the magnetic field driven case has been extensively studied in the past as well, see e.g. \cite{Walker74, Tretiakov08, Clarke08}.

\section{Model}

We study the spin texture propagation by employing the LLG equation for the magnetization ($\mathbf{S}$) dynamics with adiabatic and nonadiabatic current terms~\cite{Zhang04, Thiaville05}:
\begin{equation}
\label{eq:LLG}
\dot{\mathbf{S}}=\mathbf{S}_{0}\times 
\frac{\delta \mathcal{H}}{\delta \mathbf{S}_{0}}
-j\partial \mathbf{S}_{0}+\beta j\mathbf{S}_{0} \times \partial \mathbf{S}_{0} +\alpha \mathbf{S}_{0} \times \dot{\mathbf{S}}_{0} .
\end{equation}
Here $\mathcal{H}$ is the magnetic Hamiltonian of the system, $j$ is the
electric current in the units of velocity, $\beta$ is the
non-adiabatic spin torque constant, $\alpha$ is the Gilbert damping constant, and $\partial =\partial/\partial z$ is a derivative along the wire. We look for a solution of equation~(\ref{eq:LLG}) in the form $\mathbf{S}(z,t)
=\mathbf{S}_{0}(z,\boldsymbol\xi (t)) +\mathbf{s}$, where the
time dependence $\boldsymbol\xi (t)$ is weak, while $\mathbf{s}$ is small and
orthogonal at each point to the solution $\mathbf{S}_{0}$ of the static LLG equation.  In other words, the spin texture dynamics due to an electric current or other perturbations can be parametrized by the time-dependent  even-dimensional vector $\boldsymbol{\xi}(t)$ corresponding to the softest modes of spin texture motion. 

The equations for $\boldsymbol{\xi}(t)$ are called the effective equations of motion. For thin ferromagnetic nanowires, the DWs are rigid topological spin textures. The slow dynamics of the DW can be described in terms of only two collective coordinates
corresponding to softest modes of motion.  These modes are the DW
position $z_{0}$ and its conjugate variable -- the tilt angle $\phi$
for the transverse DW. For the vortex DW, $\phi$ is served as the
magnetization angle defining the transverse position of the vortex in
the wire \cite{Tretiakov08, Clarke08}. Up to the leading order in small dissipation ($\alpha$ and $\beta$) and current, the equations of motion take the form
\begin{eqnarray}
\dot{z}_{0}&=& -\frac{1}{2}\frac{\partial E}{\partial \phi} +j,
\label{eq:eq_z0}
\\
\dot{\phi }&=&\frac{1}{2} \frac{\partial E}{\partial z_{0}} 
-\frac{\alpha a_{zz}}{2} \frac{\partial E}{\partial \phi}
+H +(\alpha-\beta)a_{zz}j.
\label{eq:eq}
\end{eqnarray}
Here for simplicity we take the magnetic field $\mathbf{H}$ to be along the wire
(in the $z$ direction), $a_{zz}=\frac{1}{2} \int dz (\partial_z
\mathbf{S}_0)^2$, and $E(\mathbf{\xi})=\mathcal{H}_{\delta}
[\mathbf{S}_{0}(z,\boldsymbol{\xi})]$ is the energy of the domain wall
\cite{Liu11} as a function of the soft modes $\boldsymbol{\xi}$.  These equations are universal and do not depend on details
of the microscopic model. The only required input is the energy
of a static DW as a function of $z_{0}$ and
$\phi$. The latter function can be either evaluated by means of micromagnetic simulations or an approximate analytical model,  as well as experimentally measured for a given nanowire through induced emf due to the DW dynamics \cite{Liu11}.

\section{Results and Discussion}

\subsection{Translationally noninvariant nanowires}

\begin{figure}
\begin{center}
\includegraphics[width=0.9\columnwidth]{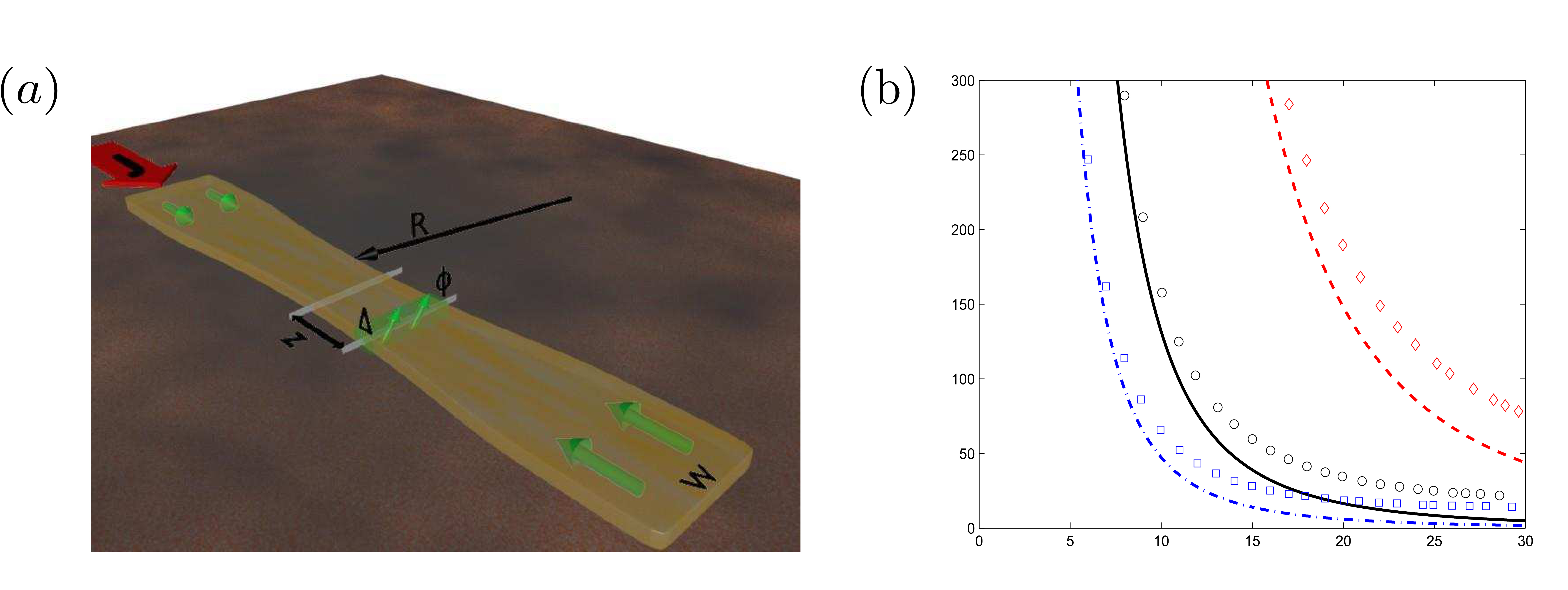}
\end{center}
\caption{(a) A sketch of a magnetic memory nanodevice based on the hourglass-shaped nanostrip. The red arrow represents the current in the $\hat{z}$ direction along the wire, the tilt angle $\phi $ constrained to the $x-y$ plane, the position $z$, and DW width $\Delta $ of the DW are shown. (b) Dependence of the switching energy in arbitrary units on the switching time.}
\label{fig:hourglass}
\end{figure}  

Equations (\ref{eq:eq_z0}) --(\ref{eq:eq}) allow the description of the DW dynamics in translationally noninvariant nanowires \cite{Tretiakov2012}. Thus one can treat the effects of DW pinning in a wire and consider nanostrips of varying width. As one of the examples of the usage of equations (\ref{eq:eq_z0}) --(\ref{eq:eq}) we consider a magnetic memory
device based on a flat hourglass-shaped nanostrip sketched in
Fig.~\ref{fig:hourglass} (a). We propose a
nonvolatile device, which employs the magnetization direction within
the DW as the information storage. Without applied current, a transverse DW
stays at the narrowest place in the nanostrip. When a certain current pulse is applied, the DW
magnetization angle $\phi$ flips from $0$ to $\pi$. At the
intermediate step of this switching process, the DW also deviates from
the narrowest part of the nanostrip but at the end it comes back
with the opposite magnetization direction in the center of the DW.  At a later time, the same current pulse can move it back to the original configuration.

The time that it takes to switch the magnetization depends on the current pulse
shape. During this process the main energy loss in a realistic wire is the
Ohmic loss. How much energy is needed for a single switching also
depends on the parameters of the current pulse. Using equations
(\ref{eq:eq_z0}) --(\ref{eq:eq}) we find the
optimal current pulse shape for
a given switching time and the minimal required energy per flip as a function
of the switching time, Fig.~\ref{fig:hourglass} (b) \cite{Tretiakov2012}.

\subsection{Ferromagnetic nanowires with Dzyaloshinskii-Moriya interaction}

\begin{figure}
\begin{center}
\includegraphics[width=0.5\columnwidth]{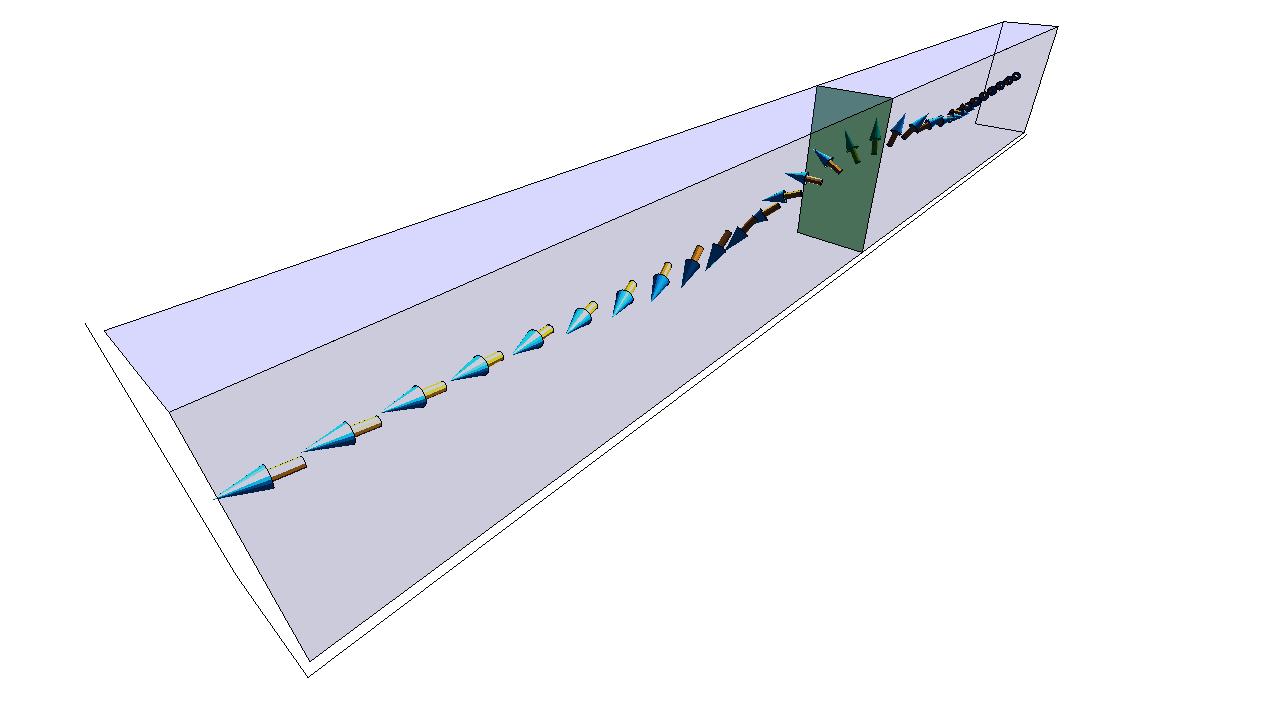}
\end{center}
\caption{A sketch of a spin-spiral domain wall in a nanowire with DM interaction.}
\label{fig:spiralDMI}
\end{figure}  

One can also use Eqs.~(\ref{eq:eq_z0}) --(\ref{eq:eq}) to describe the ferromagnetic nanowires with Dzyaloshinskii-Moriya interaction (DMI) \cite{Tretiakov_DMI}. Recently  the spiral structure of
magnetization due to DMI has been experimentally observed \cite{Uchida06, Ferriani08, Meckler09}. To describe the spin spiral DWs in ferromagnetic nanowires one should use the Hamiltonian:
\begin{equation}
\mathcal{H}=\int
dz\left[\frac{J}{2}\left(\partial_z\mathbf{S}\right)^{2}
+D\mathbf{S}\cdot\left(\mathbf{e}_{z}\times
\partial_z\mathbf{S}\right)-\lambda S_z^{2} \right].
\label{eq:Ham}
\end{equation}
Here $\mathbf{S}$ is the normalized
magnetization vector, $J>0$ is exchange interaction constant, $D$ is the DMI constant assuming that
the wire is cut or grown along the DMI vector. We consider a  thin uniform ferromagnetic wire which can be modeled as a
one-dimensional classical spin chain, where the wire is along the $z$-axis.  The last term in
Eq.~\eqref{eq:Ham} is due to uniaxial anisotropy (with the anisotropy
constant $\lambda$) which shows that the system
favors the magnetization along the wire. A transverse anisotropy is also added later as a perturbation. By minimizing the Hamiltonian~(\ref{eq:Ham}) with the appropriate boundary conditions one can get a spin spiral DW. Using the equations of motion~(\ref{eq:eq_z0}) --(\ref{eq:eq}) with the Hamiltonian~(\ref{eq:Ham}), one obtains \cite{Tretiakov_DMI}:
\begin{eqnarray}
\dot{z}_{0} &=& \frac{\beta}{\alpha}j
+\frac{(\alpha-\beta)(1+\alpha\Gamma\Delta)}{\alpha(1+\alpha^{2})}
\left[j-j_{c}\sin(2\phi)\right],
\label{eq:AdotZ0}\\
\dot{\phi} &=&\frac{(\alpha-\beta)
\Delta}{(1+\alpha^{2})\Delta_{0}^{2}}\left[j-j_{c}\sin(2\phi)\right],
\label{eq:AdotPhi}
\end{eqnarray}
where $j_{c}$ is the critical current above which the spin spiral starts rotating around the axis of the nanowire, $\Delta$ is the DW width ($\Delta_0$ is the DW width in the absence of DMI), and  $\Gamma=D/J$ is the pitch of the spin spiral. A snapshot of a moving spin-spiral DW in a nanowire with DMI is shown in Fig.~\ref{fig:spiralDMI}. Based on these equations one can study the influence of DMI on the critical current and drift velocity of domain wall \cite{Tretiakov_DMI}.

\subsection{Time-dependent currents and Ohmic losses}

For the highest performance of DW memory or logic devices, it is important to minimize the Ohmic losses in the wire, which are due to the resistance of the wire
itself and the entire circuit. They are proportional to the
time-averaged current square, $\langle j^2 \rangle$. Their
minimization has a twofold advantage. First, one can increase the
maximum current which still does not destroy the wire by excessive
heating and therefore move the DWs with a higher velocity, since the
DW velocity increases with the applied current.  Second, it creates
the most energy efficient memory devices and increases their
reliability.

These goals can be achieved by utilizing ``resonant''
time-dependent current pulses, which allow to gain a significant reduction
of Ohmic losses.  Based on the DW equations of motion, we show in the next section that all thin wires can be characterized by
three parameters obtained from dc-driven DW motion experiments:
critical current $j_{c}$, drift velocity $V_{c}$ at the critical current, and material dependent parameter $a>0$, which in particular
depends on Gilbert damping $\alpha$ and non-adiabatic spin torque
constant $\beta$. The parameter $a$ is just a ratio of the slopes of
the drift velocity $V_{d}(j)$ at large and small dc-currents \cite{Tretiakov:losses}.  We find the minimal power
$\langle j^2 \rangle$ needed to drive a DW with drift velocity $V_d$.
We show that there is a significant reduction in the heating power compared
to dc current above certain DW drift velocity and the exact
time-dependence of the optimal current pulses is found \cite{Tretiakov:losses, Tretiakov:JAP}.

\subsection{Voltage induced by moving spin textures}

\begin{figure}
\begin{center}
\includegraphics[width=0.45\columnwidth]{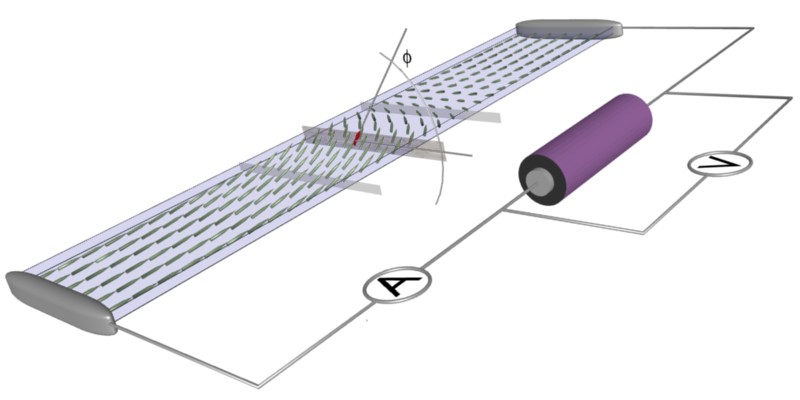}
\end{center}
\caption{A domain wall moved by a constant DC current generates a time-dependent voltage which, if measured, can give a knowledge of the DW dynamics.}
\label{fig:spinEMF}
\end{figure}

For the uniform wires without pinning, Eqs.~(\ref{eq:eq_z0}) --(\ref{eq:eq}) can be reduced to
\begin{eqnarray}
\label{z}
&&\dot{z}_0=Aj+B[j-j_c\sin(2\phi)],\\
\label{phi}
&&\dot{\phi}=C[j-j_c \sin(2\phi)].
\end{eqnarray}
The coefficients $A$, $B$, $C$ and the critical current $j_c$ are the
parameters that fully describe the DW dynamics. Using these equations one can determine the DW dynamics by measuring the voltage induced by moving DW in a uniform nanowire \cite{Liu11}, a sketch of such measurement scheme is shown in Fig.~\ref{fig:spinEMF}.

In general, the DW energy has two contributions: the power supplied by an
electric current and a negative contribution due to dissipation in the
wire.  Using the general solution of the LLG, Eq.~\eqref{eq:LLG}, one
can obtain the derivative of the energy as 
\begin{eqnarray}
\label{E_dot}
\dot{E} = 2[\beta a_{zz} \dot{z}_{0} +(1 -\beta  a_{z\phi }) \dot{\phi }]j
- \alpha \int dz \dot{\mathbf{S}}_0^2 ,
\end{eqnarray}
where $a_{z\phi}= \frac{1}{2} \int dz \partial_z \mathbf{S}_0\cdot
\partial_{\phi} \mathbf{S}_0$. The last term on the right-hand side of Eq.~\eqref{E_dot} describes
the dissipation and therefore is always negative. Meanwhile, the
first term is proportional to the current $j$ and gives the power
$Vj$ supplied by the current. We then can obtain the expression for the
induced DW voltage,
\begin{equation}
\label{voltage}
V=\frac{A^2C}{B}j +C(1+A)[j-j_c \sin (2\phi)].
\end{equation}

Equation~\eqref{voltage} describes the contribution to the voltage
due to spin-texture dynamics. This contribution is in addition to the usual Ohmic
one.  The voltage $V$ in Eq.~\eqref{voltage} is measured in units of
$Pg\mu_B/(e\gamma_0)$, where $P$ is the current polarization. We emphasize
that unlike in the previously studied cases \cite{BeachPRL09,
  Yang2010}, this voltage is not caused by the motion of topological
defects (vortices) transverse to the wire. Based on Eq.~(\ref{voltage}) we propose a set of independent measurements of the induced emf by a moving DW \cite{Liu11}.  Furthermore, the proposed systematic approach can be used to compare the extracted
phenomenological parameters of the DW dynamics for a system described
by arbitrary underlying Hamiltonian to those of microscopic theories.

\subsection{Extension to DW dynamics in antiferromagnets}

To conclude we note that a similar collective coordinate approach can be extended to antiferromagnets to study the spin texture dynamics.  In antiferromagnets (AFMs), the dynamics is more complex because of the coupling between the staggered field and magnetization. Nevertheless, using appropriate collective coordinates one can be able to describe the AFM dynamics and show that it is equivalent to the motion of a massive particle subjected to friction and external forces. One also finds that in AFMs the currents induce DW motion by means of dissipative rather than reactive torques \cite{Tveten2013}.

\section{Summary}

We have shown that the topological spin texture dynamics in ferromagnetic nanostrips and wires obeys universal equations of motion. These equations can include the effects of  Dzyaloshinskii-Moriya interaction and thus describe the motion of spin-spiral domain walls. Based on the collective coordinate approach, we can also describe the DW dynamics due to time-dependent currents whose optimization greatly reduces Ohmic losses in the nanowires. This approach allows for the treatment of translationally noninvariant magnetic wires, thus taking into account the DW pinning and leading to the description of novel spintronic nanodevices. 

\section{Acknowledgments}

We acknowledge support  by the Grants-in-Aid for Scientific Research (No. 25800184 and No. 25247056) from the MEXT, Japan, the Welch Foundation (A-1678), and the SpinNet.


\bibliography{magnetizationDynamics}

\end{document}